\documentclass[twocolumn,superscriptaddress,showpacs,preprintnumbers,amsmath,amssymb]{revtex4}

\usepackage{graphicx,color}
\usepackage{epsfig}
\usepackage{dcolumn}
\usepackage{bm}

\begin{document}

\title{Spin-orbit and orbit-orbit strengths for radioactive neutron-rich doubly magic nucleus $^{132}$Sn
in relativistic mean field theory}

\author{Haozhao Liang}
 \affiliation{State Key Laboratory of Nuclear Physics and Technology, School of Physics,
Peking University, Beijing 100871, China}

\author{Pengwei Zhao}
 \affiliation{State Key Laboratory of Nuclear Physics and Technology, School of Physics,
Peking University, Beijing 100871, China}

\author{Lulu Li}
 \affiliation{State Key Laboratory of Nuclear Physics and Technology, School of Physics,
Peking University, Beijing 100871, China}

\author{Jie Meng}
\affiliation{School of Physics and Nuclear Energy Engineering, Beihang University,
              Beijing 100191, China}
 \affiliation{State Key Laboratory of Nuclear Physics and Technology, School of Physics,
Peking University, Beijing 100871, China}
 \affiliation{Department of Physics, University of Stellenbosch, Stellenbosch, South Africa}

\date{\today}

\begin{abstract}
Relativistic mean field (RMF) theory is applied to investigate the
properties of the radioactive neutron-rich doubly magic nucleus
$^{132}$Sn and the corresponding isotopes and isotones. The
two-neutron and two-proton separation energies are well reproduced
by the RMF theory. In particular, the RMF results agree with the
experimental single-particle spectrum in $^{132}$Sn as well as the
Nilsson spin-orbit parameter $C$ and orbit-orbit parameter $D$ thus
extracted, but remarkably differ from the traditional Nilsson
parameters. Furthermore, the present results provide a guideline for
the isospin dependence of the Nilsson parameters.
\end{abstract}

\maketitle


The concept of magic numbers is one of the most fundamental
ingredients for understanding the nature of atomic nuclei. Due to
the strong spin-orbit couplings, the magic numbers for stable nuclei
are shown as 2, 8, 20, 28, 50, and 82 for both protons and neutrons
as well as 126 for neutrons \cite{Haxel1949,Goeppert-Mayer1949},
which are no longer simply the shell-closure of the harmonic
oscillators. Thus, it is quite sophisticated to predict the next
proton and neutron magic numbers \cite{Zhang2005,Meng2006}, which
are, nevertheless, critical for guiding the superheavy element
synthesis. The phenomenon of shell-closure is also crucial for
determining the waiting-points of the rapid neutron-capture process
($r$-process), which is responsible for the production of more than
half of the elements heavier than iron.

With both proton and neutron magic numbers, the so-called doubly
magic nuclei form a very small and exclusive club. These nuclei are
rigidly spherical and particularly stable compared to their
neighbors. Along the $\beta$ stability line, only five nuclei are
included, i.e., $^4$He, $^{16}$O, $^{40}$Ca, $^{48}$Ca, and
$^{208}$Pb. Furthermore, by simply combining these traditional magic
numbers, one will also end up with the neutron-deficient nuclei
$^{48}$Ni, $^{56}$Ni, and $^{100}$Sn, neutron-rich nuclei $^{78}$Ni
and $^{132}$Sn, as well as the extreme neutron-rich nucleus
$^{70}$Ca that is predicted as a giant halo nucleus \cite{Meng2002}.
It has been shown that the shell structures existing in the
single-particle spectra can change in the nuclei far away from the
stability line \cite{Sorlin2008}. For example, the $N=28$
shell-closure disappears when the proton number decreases, and thus
the nucleus $^{44}$S is found to have prolate-spherical shape
coexistence \cite{Force2010} and $^{42}$Si is well deformed
\cite{Bastin2007}. Therefore, whether the potentially doubly magic
nuclei far away from the stability line are indeed doubly magic is a
hot and fundamental topic.

Since $^{56}$Ni was confirmed as a radioactive doubly magic nucleus
on the neutron-deficient side \cite{Rehm1998}, the potentially
doubly magic nucleus $^{132}$Sn on the neutron-rich side has been
paid much attention both experimentally and theoretically in the
past years. However, pinning down whether $^{132}$Sn is doubly magic
or not was not so straight forward. By using the Sn($\alpha$,$t$)
reactions, it was found that outside the $Z=50$ core the energy gap
between the proton single-particle states $1h_{11/2}$ and $1g_{7/2}$
increases with neutron excess \cite{Schiffer2004}, which suggests a
decrease in the nuclear spin-orbit interaction. This was reported in
\textit{Nature} in an article entitled \textit{Not-so-magic numbers}
\cite{Warner2004}. Very recently, by using the
$^{132}$Sn($d$,$p$)$^{133}$Sn reaction, the experiment performed at
Oak Ridge National Laboratory revealed for the first time that the
spectroscopic factors of the neutron single-particle states
$3p_{1/2}$, $3p_{3/2}$, $2f_{5/2}$, and $2f_{7/2}$ outside the
$N=82$ core are consistent with $S\approx1$. This is one of the
critical pieces of evidence to conclude that $^{132}$Sn is perfectly
doubly magic, even better than $^{208}$Pb, which has attracted vast
attention \cite{Cottle2010,Schwarzschild2010}.

\begin{figure}
\includegraphics[width=0.45\textwidth]{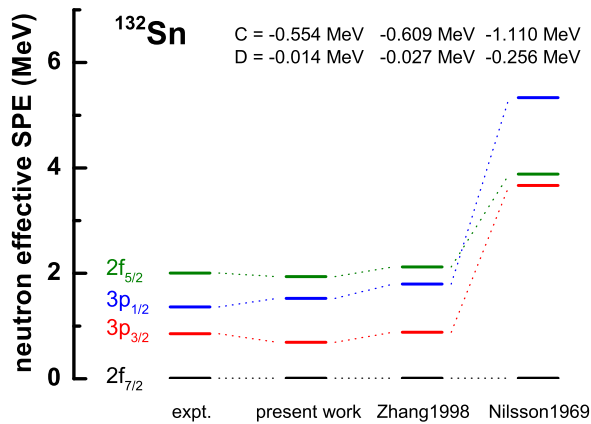}
\caption{(Color online) Neutron effective single-particle energies (SPE)
    with respect to the orbital $2f_{7/2}$ in $^{132}$Sn.
    From left to right, the experimental SPE \cite{Jones2010}
    and those calculated by
    the Nilsson model with spin-orbit $C$ and orbit-orbit $D$ parameters re-fitted in the present work
    as well as the parameters given in Ref.~\cite{Zhang1998} and the traditional ones \cite{Nilsson1969}
    are shown.
    \label{Fig1}}
\end{figure}

The single-particle states outside a hard double-closure core are
crucial for calibrating the theoretical models, especially pure mean
field theories, since these valance states are well isolated from
the core, and not fragmented. As the only confirmed neutron-rich
doubly magic nucleus so far, $^{132}$Sn does show unique
characteristics in the single-particle spectra. In Fig.~\ref{Fig1},
the neutron effective single-particle energies (SPE) with respect to
the orbital $2f_{7/2}$ are shown. The experimental SPE
\cite{Jones2010} shown in the left row are compared with those
calculated by the Nilsson model with traditional spin-orbit and
orbit-orbit parameters \cite{Nilsson1969} shown in the right row. It
can be clearly seen that, even though the Nilsson model has achieved
great success in describing the single-particle structure for stable
nuclei, the traditional parameters encounter a serious problem in
reproducing the single-particle spectrum of such neutron-rich
nucleus. In other words, the single-particle spectra of exotic
nuclei are quite different from those of stable nuclei. Zhang
\textit{et al.} pointed out that the Nilsson spin-orbit parameter
$C=-2\kappa\hbar\omega$ and the orbit-orbit parameter
$D=-\kappa\mu\hbar\omega$ should be re-fitted in this case
\cite{Zhang1998}. Since the experimental SPE has been changed
compared to those used in Ref.~\cite{Zhang1998}, the parameters $C$
and $D$ are re-fitted according to the latest experimental SPE in
$^{132}$Sn in the present work, as shown in Fig.~\ref{Fig1}. It
turns out that the re-fitted spin-orbit parameter $C$ reduces
$50\%$, and the orbit-orbit parameter $D$ is of one order of
magnitude smaller than the traditional one.

The relativistic mean field (RMF) theory \cite{Meng2006}, which has
received wide attention due to its successful description of a large
variety of nuclear phenomena during the past three decades, is
regarded as one of the best microscopic candidates for the
descriptions of both stable and exotic nuclei. In this framework,
the combination of the scalar and vector fields, which are of the
order of a few hundred MeV, provide a natural and more efficient
description of both the nuclear mean field central and spin-orbit
potentials. So far, the most widely used RMF framework is based on
meson-exchange or point-coupling effective interactions. In each
case, a quantitative treatment of nuclear matter and finite nuclei
needs a medium dependence of effective mean-field interactions,
which can be taken into account by the inclusion of higher-order
(nonlinear coupling) interaction terms or by assuming a density
dependence for the coupling interactions.

In this paper, various versions of the RMF theory will be applied to
investigate the properties of the neutron-rich doubly magic nucleus
$^{132}$Sn and the corresponding isotopes and isotones. The
two-neutron and two-proton separation energies, the single-particle
spectrum, as well as the spin-orbit and orbit-orbit strengths thus
extracted will be discussed.


\begin{figure}
\includegraphics[width=0.45\textwidth]{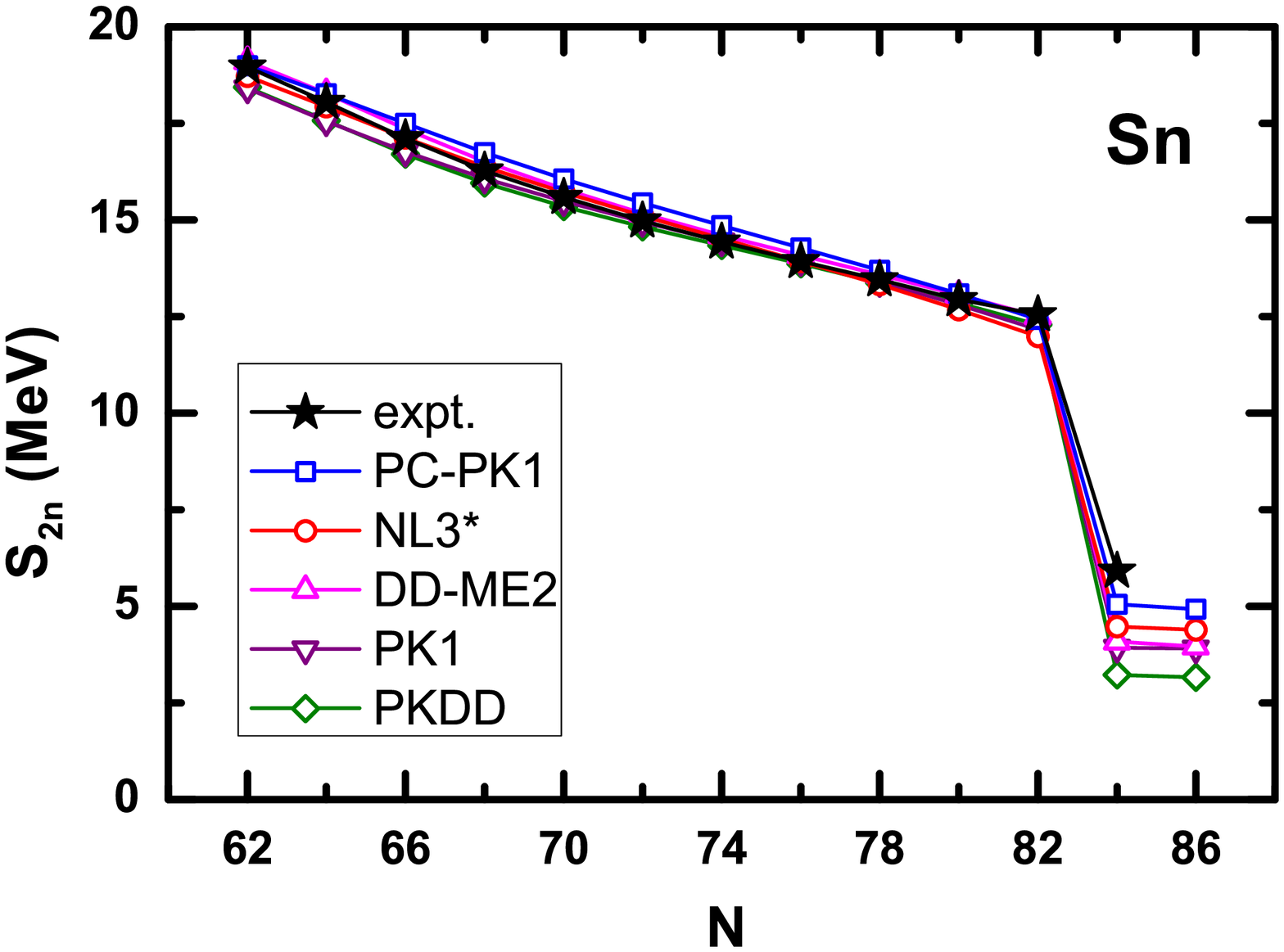}\\
\includegraphics[width=0.45\textwidth]{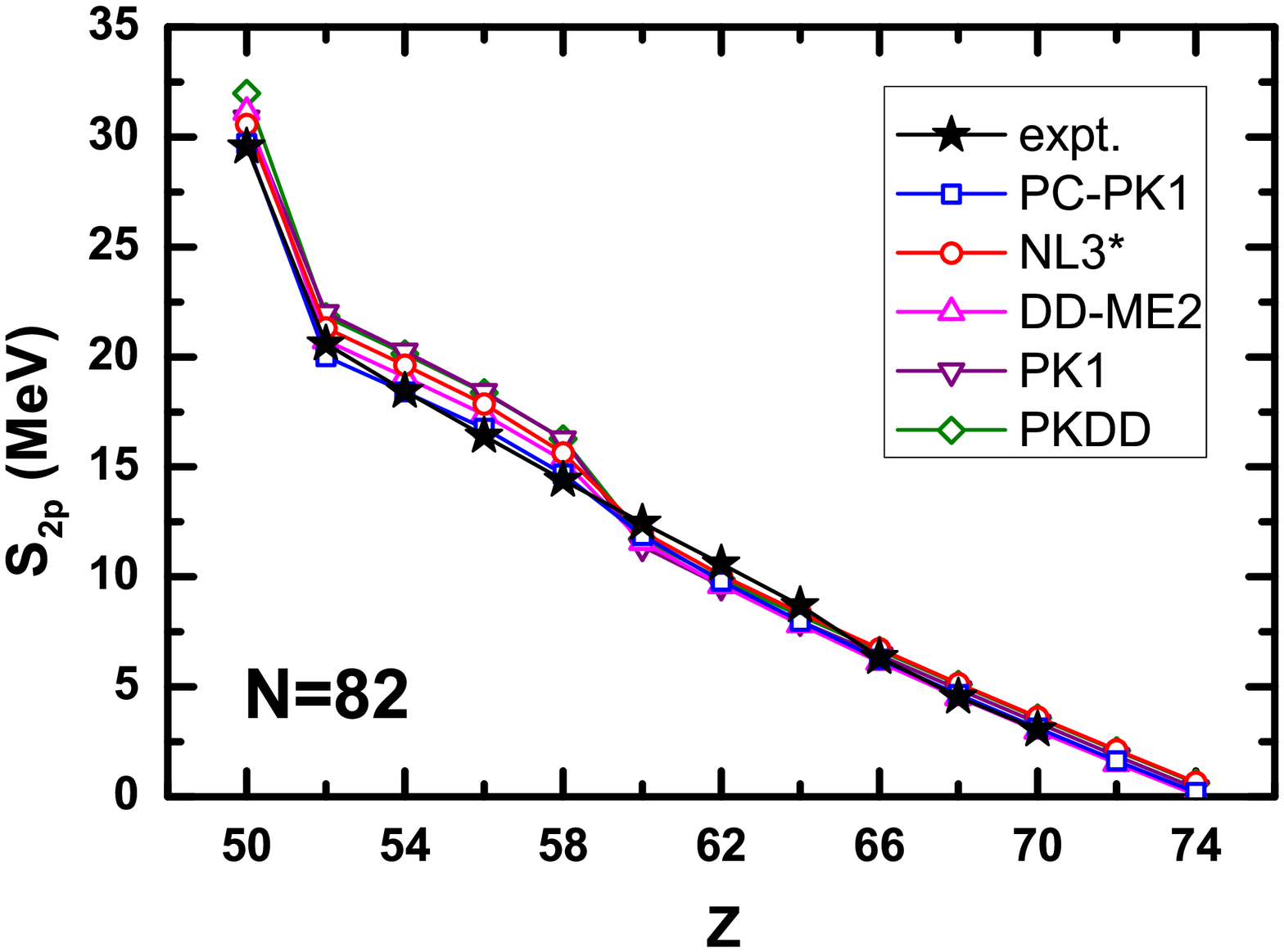}
\caption{(Color online) Two-neutron (upper panel) and two-proton
    (lower panel) separation energies
    of the Sn isotopes and $N=82$ isotones calculated by the RMF theory with
    the effective interactions PC-PK1 \cite{Zhao2010}, NL3* \cite{Lalazissis2009}, DD-ME2 \cite{Lalazissis2005},
    PK1 \cite{Long2004}, and PKDD \cite{Long2004}.
    The mass data of $^{132}$Sn and $^{134}$Sn are taken from Ref.~\cite{Dworschak2008},
    and the others from Ref.~\cite{Audi2003}.
    \label{Fig2}}
\end{figure}

In Fig.~\ref{Fig2}, the two-neutron separation energies $S_{2n}$ of
the Sn isotopes and the two-proton separation energies $S_{2p}$ of
the $N=82$ isotones are shown in the upper and lower panels,
respectively. The mass data of $^{132}$Sn and $^{134}$Sn are taken
from Ref.~\cite{Dworschak2008} and the others from
Ref.~\cite{Audi2003}. The theoretical results are calculated by the
RMF theory with non-linear point-coupling effective interaction
PC-PK1 \cite{Zhao2010}, and non-linear finite-range interactions
NL3* \cite{Lalazissis2009} and PK1 \cite{Long2004}, as well as
density-dependent finite-range interactions DD-ME2
\cite{Lalazissis2005} and PKDD \cite{Long2004}. It is shown that the
experimental data can be well reproduced. In particular, all
theoretical results show profound jumps in $S_{2n}$ and $S_{2p}$ at
$N=82$ and $Z=50$, which indicates that $^{132}$Sn is predicted as a
legitimate doubly magic nucleus by all the effective interactions
used here. It is also found that the size of the neutron shell gaps
are slightly overestimated due to the relatively low effective mass.
Another critical test of the shell-closure is to study
single-particle spectra, especially the single-particle states
outside the rigidly spherical core, which can be measured precisely
in experiment.

\begin{figure*}
\includegraphics[width=0.60\textwidth]{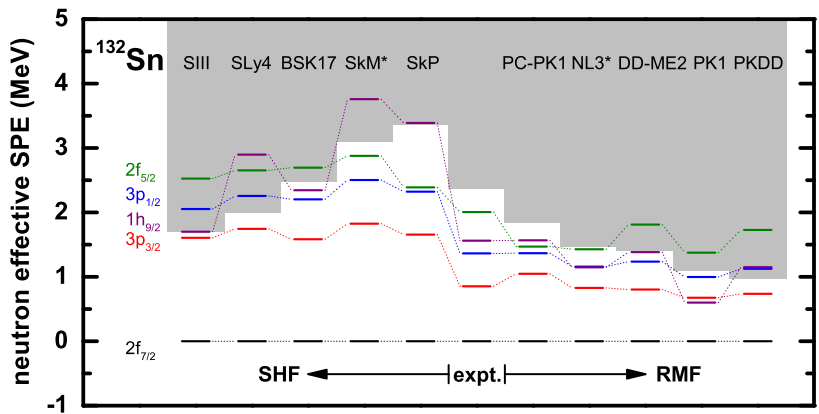}
\caption{(Color online) Neutron effective SPE
    with respect to the orbital $2f_{7/2}$ in $^{132}$Sn calculated by the RMF theory with
    effective interactions PC-PK1, NL3*, DD-ME2, PK1, and PKDD.
    The experimental data \cite{Jones2010} and the results calculated by the Skyrme-Hartree-Fock
    (SHF) theory with effective interactions SkP~\cite{Dobaczewski1984}, SkM*~\cite{Bartel1982},
    BSK17~\cite{Goriely2009}, SLy4~\cite{Chabanat1998}, and SIII~\cite{Beiner1975} are also shown for comparison.
    The shaded areas indicate the area beyond the neutron threshold.
    \label{Fig3}}
\end{figure*}

In Fig.~\ref{Fig3}, the neutron effective SPE with respect to the
orbital $2f_{7/2}$ in $^{132}$Sn calculated by the RMF theory with
effective interactions PC-PK1, NL3*, DD-ME2, PK1, and PKDD are
compared with the experimental data \cite{Jones2010}, where the
shaded areas indicate the area beyond the neutron threshold. The
results calculated by the Skyrme-Hartree-Fock (SHF) theory with
typical effective interactions SkP~\cite{Dobaczewski1984},
SkM*~\cite{Bartel1982}, BSK17~\cite{Goriely2009},
SLy4~\cite{Chabanat1998} and SIII~\cite{Beiner1975} are also shown
for comparison. It should be noted that the overflow of some
single-particle states beyond the neutron threshold results from the
low effective nucleon mass. An additional enhancement of the
effective mass in finite nuclei can be caused by the coupling of
single-nucleon levels to low-energy collective vibrational states,
which is an effect entirely beyond mean field. In the present
mean-field level, it is clearly shown that the overall structure of
the neutron effective SPE for such a neutron-rich doubly magic
nucleus can be well reproduced in the relativistic framework, where
both the spin-orbit and orbit-orbit potentials are described in a
self-consistent way. In contrast, the non-relativistic results
overestimate nearly twice the single-particle energy spacing between
the orbitals $3p_{3/2}$ and $2f_{7/2}$, where the reduction of the
orbit-orbit potential plays the most important role.

\begin{figure}
\includegraphics[width=0.45\textwidth]{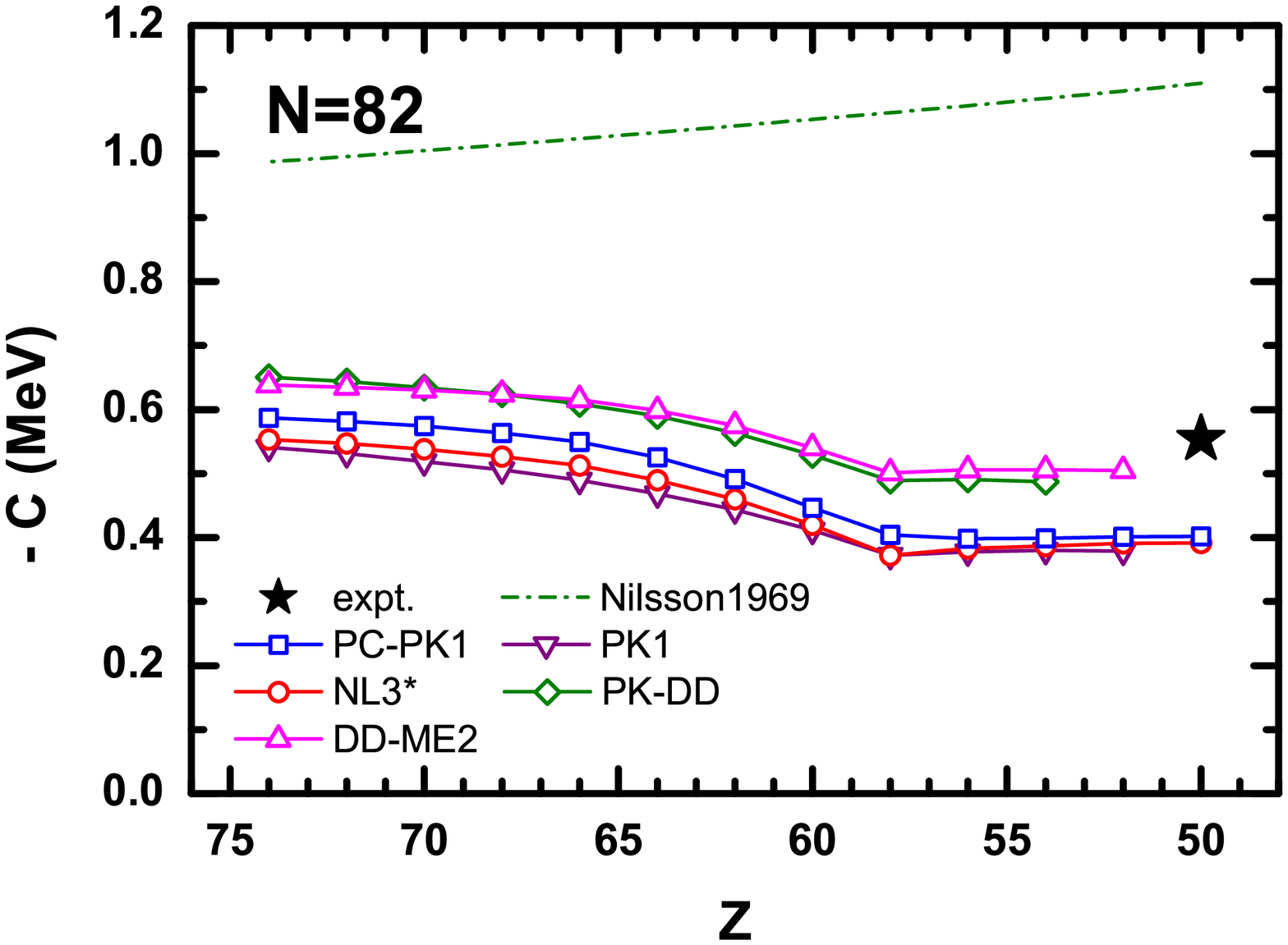}\\
\includegraphics[width=0.45\textwidth]{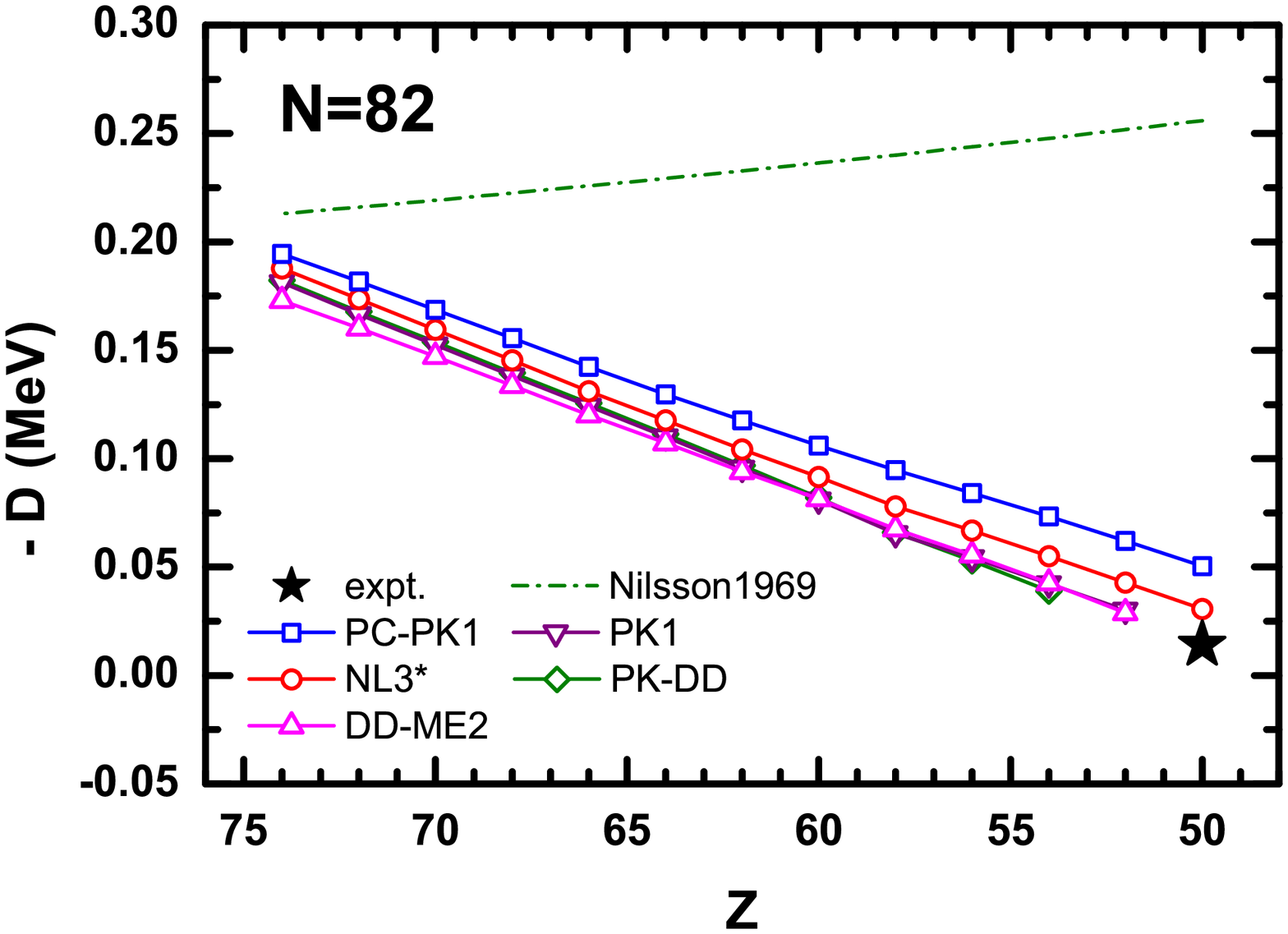}
\caption{(Color online) Nilsson spin-orbit (upper panel) and orbit-orbit (lower panel) parameters
    for the $N=82$ isotonic chain.
    Those extracted with the experimental data and RMF calculations are labeled with solid and open symbols, respectively.
    The traditional parameters from Ref.~\cite{Nilsson1969} are shown with dash-dotted lines.
    \label{Fig4}}
\end{figure}

From the single-particle spectrum, one can extract the Nilsson
parameters by fitting to the single-particle energies. The
spin-orbit strength $C$ and orbit-orbit strength $D$ are two
important characteristics in the analysis of the single-particle
spectrum. The Nilsson parameters $C$ and $D$ here are obtained by
fitting to the single-particle energies of the states $3p_{1/2}$,
$3p_{3/2}$, $2f_{5/2}$ and $2f_{7/2}$ with the least-squares method.

In order to investigate the evolutions from stable to the
neutron-rich nuclei, the Nilsson spin-orbit and orbit-orbit
parameters for the $N=82$ isotonic chain thus obtained are shown in
the upper and lower panels of Fig.~\ref{Fig4}, respectively. The
results are shown only for the nuclei in which all four states
$3p_{1/2}$, $3p_{3/2}$, $2f_{5/2}$ and $2f_{7/2}$ are bound. The
re-fitted Nilsson parameters for the latest experimental data of
$^{132}$Sn are plotted with stars. It is noticed that only the data
of $^{132}$Sn are shown here, since the fragmentation in the
single-particle states is not negligible for the nuclei that are not
doubly magic. In addition, the traditional parameters taken from
Ref.~\cite{Nilsson1969} are also shown with dash-dotted lines for
comparison.

It is shown that the absolute value of the spin-orbit strength $C$
slightly decreases with neutron excess. Its strength for the
neutron-rich doubly magic nucleus can be well reproduced by the
density-dependent effective interactions, while it is somewhat
underestimated by the non-linear interactions. The traditional
Nilsson spin-orbit strength shows a slightly escalating trend, which
deviates from the data at $^{132}$Sn significantly.

Meanwhile, the absolute value of orbit-orbit strength $D$ strength
monotonously decreases as the proton number decreases, which shows
entirely the opposite tendency of the traditional Nilsson
parameters. It is also found that the orbit-orbit strengths
extracted from RMF calculations agree quite well with the
experimental data for $^{132}$Sn, which is greatly overestimated in
comparison with the traditional Nilsson parameter.

Therefore, the spin-orbit and orbit-orbit strengths of the
neutron-rich doubly magic nucleus $^{132}$Sn can be well described
by the mean field theories within the relativistic framework.
Moreover, the reduction of both the spin-orbit and orbit-orbit
strengths along the $N=82$ isotonic chain are predicted. The
reduction shown in this neutron-rich region plays an important role
in the calculations using the Nilsson parameters, for example, the
yrast bands in $^{136}$Te and $^{142}$Xe calculated by the projected
shell model \cite{Zhang1998}.


In summary, various versions of the RMF theory have been applied to
investigate the properties of the neutron-rich doubly magic nucleus
$^{132}$Sn and the corresponding isotopes and isotones. In
particular, the single-particle spectrum as well as the spin-orbit
and orbit-orbit strengths thus extracted have been discussed in
detail. It is found that the two-neutron and two-proton separation
energies are well reproduced by the RMF theory. Moreover, the RMF
results agree quite well with the experimental single-particle
spectrum in $^{132}$Sn as well as the Nilsson spin-orbit parameter
$C$ and orbit-orbit parameter $D$ thus extracted, but remarkably
differ from the traditional Nilsson parameters. Along the $N=82$
isotonic chains, the present results show entirely different
tendency in comparison with the traditional Nilsson parameters, and
provide a guideline for the isospin dependence of the Nilsson
parameters.

This work is partly supported by State 973 Program 2007CB815000, the
NSF of China under Grant Nos. 10775004, 10947013 and 10975008, and
China Postdoctoral Science Foundation Grant No. 20100480149.

\bibliography{refSn}

\end{document}